\documentclass[aps,pra,twocolumn,superscriptaddress]{revtex4-1}
\usepackage[latin9]{inputenc}
\usepackage{amsmath}
\usepackage{amssymb}
\usepackage{mathtools}
\usepackage{graphicx}
\usepackage{braket}
\usepackage{hyperref}
\usepackage{paralist}
\usepackage{color}
\usepackage{mathrsfs}
\usepackage{amsthm}
\usepackage{centernot}
\usepackage{stmaryrd}
\usepackage{xcolor}
\usepackage{soul}

\usepackage{url}

\usepackage{breakurl}
\newtheorem*{theorem*}{Theorem}
\newtheorem*{lemma*}{Lemma}
\newtheorem*{assumption*}{Assumption}

\theoremstyle{definition}

\newcommand{\tr}[0]{\textnormal{Tr}}
\DeclareMathOperator*{\Tr}{Tr}

\DeclareMathOperator{\Texp}{\mathcal T exp}
\DeclareRobustCommand{\openzero}{\leavevmode\hbox{0\kern-.55em0}}

\newcommand{\bs}{\boldsymbol}

\newcommand{\conv}{ \rho \xmapsto[]{\mathcal L_0 + \text{CC}} \sigma} 

\makeatletter
\newcommand{\xMapsto}[2][]{\ext@arrow 0599{\Mapstofill@}{#1}{#2}}
\def\Mapstofill@{\arrowfill@{\Mapstochar\Relbar}\Relbar\Rightarrow}
\makeatother

\makeatletter

\begin{document}
\title{Mixing of quantum states under Markovian dissipation and coherent control}

\date{\today}

\author{Georgios Styliaris}
\email [e-mail address: ]{styliari@usc.edu}
\affiliation{Department of Physics and Astronomy, and Center for Quantum Information
	Science and Technology, University of Southern California, Los Angeles,
	California 90089-0484}
\author{\'Alvaro M. Alhambra}
\affiliation{Perimeter Institute for Theoretical Physics, Waterloo, ON N2L 2Y5, Canada}
\author{Paolo Zanardi}

\affiliation{Department of Physics and Astronomy, and Center for Quantum Information
Science and Technology, University of Southern California, Los Angeles,
California 90089-0484}

\begin{abstract}

Given any two quantum states $\rho$ and $\sigma$ in Hilbert spaces of equal dimension satisfying the majorization condition $\rho \succ \sigma$,
it is always possible to transform $\rho \mapsto \sigma$ by a unital quantum map. In fact, any such transformation can be achieved just by means of noisy operations, i.e., by access to maximally mixed ancillary states and unitary transformations that act jointly in the system-ancilla space. Here, we investigate the possible transitions between states (i.e., the induced preorder of states)  when one restricts the unitary control to the quantum system alone and replaces the maximally mixed ancillas with a Markovian master equation, represented by a unital Lindbladian. As a main result, we find necessary and sufficient conditions for the Lindbladian dissipation to have the same converting power  as that of noisy operations, i.e., any transformation $\rho \mapsto \sigma$ is possible if and only if $\rho \succ \sigma$. %The results \st{help us understand} \geo{are towards understanding} the resource-theoretic preorders of states when considering additional physically-motivated restrictions on the free operations \geo{of a resource theory}.\alv{I was thinking of a sentence motivating the restriction, that's an idea which may not be the best}

\end{abstract}

\maketitle

\section{Introduction}

With the advent of quantum technologies, the need to characterize the properties of quantum systems and learn how to control their dynamics becomes increasingly pressing. One of the natural frameworks within that general program is that of \emph{resource theories} \cite{chitambar2018quantum}. In it, one defines a certain restriction on a class of operations which, motivated by physical considerations, are deemed as ``easy" or ``free". For instance, in entanglement theory \cite{horodecki2009quantum}, the operations are local quantum maps and classical communication (LOCC), which appear under the natural assumption that establishing quantum channels between distant parties poses a fundamental difficulty. Starting from there, one finds a way to systematically analyze which states are more entangled than others: if $\rho$ can be transformed to $\sigma$ with LOCC operations, it is reasonable to conclude that $\rho$ is ``more entangled" than $\sigma$. These relations between states ($\rho \mapsto \sigma$ via ``easy'' operations)  induce a particular \emph{preorder} \footnote{Preorder is a binary relation that is reflexive and transitive. In a preorder ``$\ge$'', two objects $a$ and $b$ may be incomparable, i.e., it may not always hold true that either $a \ge b$ or $b \ge a$. On the other hand, $a \ge b$ and $b \ge a$ does not necessarily imply that $a = b$.}  in the space of quantum states. 

The \emph{mixedness} or \emph{uniformity} of quantum states (that is, how far they are from being in a  pure state, as opposed to a statistical ensemble) is another property that can be characterized in this way. In order to do that, the natural restrictions one imposes to the ``easy" operations are that they shall include (i) access to fully random quantum states or \emph{noise} and (ii) access to arbitrary reversible maps, i.e., unitaries. Neither of these should be able to decrease any reasonable notion of mixedness. The resource theory defined this way goes under the name of \emph{non-uniformity} or \emph{Noisy Operations} (NO) \cite{gour2015resource}, and includes all maps of the form 
\begin{equation}\label{eq:NO}
\rho \mapsto \tr_{E}\left[U\left( \rho \otimes \frac{I_E}{d_E} \right) U^\dagger \right],
\end{equation}
where $U$ is any unitary and the environment is a maximally mixed state of any dimension $d_E$ \footnote{In fact, the set of NO is defined to also include all linear maps that can be arbitrarily well approximated by \eqref{eq:NO}}. If there exists such a map with which the transition $\rho \mapsto \sigma$ is possible, we are confident in stating that ``$\sigma$ is more mixed than $\rho$". As an example, the state $I_d/d$ is the only state that is left unchanged by this set of maps, which is consistent with it being the most mixed. The mathematical framework that characterizes the preorder induced by state transitions under NO is that of \emph{majorization} of the probability distributions of the quantum state eigenvalues.

The theory of non-uniformity is part of the larger landscape of \emph{thermodynamic resource theories} \cite{horodecki2003reversible,Zyczkowski2004,brandao2013resource,sparaciari2017resource}, which are used to formalize the out of equilibrium physics of small quantum systems from a quantum-informational perspective. The aim of this general framework is to understand the different roles that energy, coherence and purity play in statistical and thermodynamical phenomena.
One of the potential limitations is that the set of ``easy" or allowed operations may still be too large, as it may include a large number of processes that are still infeasible in practice, such as arbitrarily strong interactions with a very large environment. 

Given that, it is not always clear how much physical content one should assign to the preorders of states. To address this limitation, one can ask the question: In which meaningful ways can we modify the set of operations without affecting the preorder generated by them?
This has been addressed in various ways both in the context of the resource theory of NO, as well as thermal operations (TO). The restrictions include constraints on the size of the environment \cite{sparaciari2017resource,scharlau2018quantum,richens2018finite,boes2018catalytic2} or on the interactions allowed \cite{perry2015sufficient,wilming2016second,lostaglio2018elementary,baumer2017partial}. 

In this paper, we focus on a different physically-motivated modification of the set of ``free'' operations: we assume that the environment is a single Markovian source of noise for the system and that thus the system only undergoes combinations of coherent control and unital Markovian dissipation \footnote{We note, however, that depending on the master equation the allowed set may not be within NO, as there are examples of unital semigroups which can be shown do not belong to the set \cite{haagerup2011factorization}.}. This choice is far from arbitrary: small systems are very often weakly coupled to a very large environment that is out of the reach for the experimentalist. The information that is leaked to such environments tends to dissipate very fast, in which case the noise induced in the system is well approximated by a Markovian master equation.

As a main result, we characterize the set of Markovian master equations which allow for the preservation of the majorization preorder between states. That is, the individual master equations that, together with coherent control, allow for the same state transformations as the whole set of maps of the form \eqref{eq:NO}. With this, we show that majorization captures the notion of mixedness for this class of open system dynamics. This result differs from what happens in the resource theory of TO, where restricting to Markovian semigroups already narrows quite heavily the set of state transitions allowed on a qubit \cite{lostaglio2018elementary}. We also briefly explore how other master equations affect the preorder, and speculate on the effect of different restrictions on the coherent control.
	
The power of dissipation with coherent control has already been explored in a number of previous works, with more restricted settings (see for instance \cite{schirmer2002criteria,altafini2003controllability,altafini2004coherent,
wu2007controllability,dirr2009lie,kurniawan2009controllability,o2012illustrating,yuan2012reachable,rooney2018steering}). While in most of the literature the main tools employed are Lie-algebraic, here we exploit basic results from the theory of majorization.

The paper is organized as follows. We begin in \autoref{section_Preliminaries} with the relevant preliminary background. Then, in \autoref{section_tools} we define the state convertibility problem and introduce the basic tools used later for its analysis. Section \ref{section_main} contains the main results and their derivations. We summarize and further discuss them in \autoref{section_results}.

\section{Preliminaries} \label{section_Preliminaries}

\subsection{The majorization preorder of probability distributions}

The notion of how mixed different distributions are is well captured by the theory of majorization.
Let $\bs p = \left(p_1,p_2 , \dotso ,p_d \right)$ be a vector representing a discrete probability distribution of $d$ possible outcomes. One can pick two components $p_i$ and $p_j$ of $\bs p$  and combine them, producing two new components $p'_i$ and $p'_j$, using the mixing rule
\begin{align} \label{T_transform_example}
\begin{pmatrix}
p'_i \\
p'_j
\end{pmatrix}
= \begin{pmatrix}
1-s & s \\
s & 1-s
\end{pmatrix}
\begin{pmatrix}
p_i \\
p_j
\end{pmatrix}
\end{align}
for some $s \in \left[ 0 , 1  \right]$.  The resulting probability vector is $\bs p' = \left( p_1,p_2, \dotso, p'_i, \dotso, p'_j,\dotso,p_d \right)$. This mixing of two levels is known as \emph{T-transform}. One can apply a sequence of transformations of the form \eqref{T_transform_example}, possibly between a different pairs of components each time. After each round, the probability distribution becomes increasingly \textit{mixed}.

%In other words, a source of ``randomness'' is necessary for mixing.

%What does it mean to mix? Let $\bs p = \left(p_1,p_2 , \dotso ,p_d \right)$ be a vector representing a discrete probability distribution of $d$ outcomes. Then, one can pick two components of $\bs p$  (say $p_i$ and $p_j$) and combine them producing two new components (say $p'_i$ and $p'_j$) using the blending rule
%\begin{align} \label{T_transform_example}
%\begin{pmatrix}
%p'_i \\
%p'_j
%\end{pmatrix}
%= \begin{pmatrix}
%1-s & s \\
%s & 1-s
%\end{pmatrix}
%\begin{pmatrix}
%p_i \\
%p_j
%\end{pmatrix}
%\end{align}
%for some $s \in \left[ 0 , 1  \right]$. This two-level blending is known as a T-transform \cite{bengtsson2017geometry}. The resulting vector $\bs p' = \left( p_1,p_2, \dotso, p'_i, \dotso, p'_j,\dotso,p_d \right)$ is still a probability vector, representing a ``more uniform'' distribution. One can keep scrambling the components, with each round producing a more and more mixed probability vector.
One can also think of more general $d \times d$ \textit{stochastic} matrices ($M_{ij}$ with non-negative entries such that $\sum_i M_{ij} = 1$ for all $j$), i.e., matrices that map probability vectors to probability vectors. The relevant subset of stochastic matrices are called \textit{bistochastic} (or \textit{doubly-stochastic}), which are those whose matrix elements satisfy the additional condition $\sum_j (M)_{ij} = 1$ for all $i$. This is equivalent to having the maximally mixed probability vector $\frac{1}{d} \left(1,1,\dotso,1\right)$ as a fixed point. For $d>2$, the set of sequences of T-transforms is a strict subset of bistochastic matrices, while for $d=2$ is coincides with it.

Even though one is a subset of the other, the possible transitions between probability distributions that they allow for are, in fact, the same. This is summarized in the following theorem, central to the theory of majorization:
\begin{theorem*} Let $\bs{p}=(p_1,\dots,p_d)$ and $\bs{p}'=(p'_1,\dots,p'_d)$ be probability distributions. The following statements are equivalent:
	\begin{enumerate}[(i)]
		\item Let $(p^\downarrow_1,\dots,p^\downarrow_d)$ be a permutation of $\bs{p}$ such that $p^\downarrow_i\ge p^\downarrow_{i+1}$, and similarly for $\bs{p}'$. Then, for every $k \in \{1,\dotso,d\}$,
		\begin{equation}
		\sum_{i=1}^k p^\downarrow_i  \ge \sum_{i=1}^k p'^{\downarrow}_i.
		\end{equation} 
		\item There exists a bistochastic matrix $M$ such that $\bs{p}'=M \bs{p}$.
		\item There exists a sequence of at most $(d-1)$ T-transforms $T_i$ such that $\bs{p}'= \left( \prod_i T_i \right) \bs{p}$
	\end{enumerate}
\end{theorem*}
If $\bs{p}$ and $\bs{p'}$ are such that these conditions hold, we say that $\bs{p}$ \emph{majorizes} $\bs{p'}$, which we denote by $\bs p \succ \bs p'$. It is reasonable to conclude that $\bs p'$ being ``more mixed" than $\bs p$ is precisely captured by the statement ``$\bs p \succ \bs p'$". For more details on the theory of majorization and for a proof of the theorem above, we refer the reader to \cite{marshall2010inequalities,nielsen2001majorization}.

\subsection{Mixing of quantum states}
%\alv{how do you want to write this?}
%In the quantum formalism, states are represented by density matrices, which are positive semidefinite matrices with unit trace. \st{As such, any state can be written in the form $\rho=\sum_i \lambda^{(\rho)}_i \ket{i}\bra{i}$ where $\{ \lambda^{(\rho)}_i \}$ is a probability distribution and $\{ \ket{i} \}$ forms an orthonormal basis. This way, one can understand that} a quantum state is ``more mixed" the more uniform their eigenvalues are. \geo{I think spectral theorem is too much. We can discuss.}

In the quantum formalism states are represented by density matrices, which are positive semidefinite matrices with unit trace, so their eigenvalues form a probability distribution. A quantum state is ``more mixed" the more uniform the probability distribution of its eigenvalues is.  We denote $\rho \succ \sigma$ to indicate the majorization of the corresponding probability distributions of the spectra.
%, which we write as $\bs{\lambda}^{(\rho)} \succ  \bs{\lambda}^{(\sigma)}$.

The meaning of ``more mixed" is justified by thinking back at the definition of NO in Eq. \eqref{eq:NO}. The central result of the theory is that there exists a NO taking $ \rho \xmapsto[]{NO} \sigma$ if and only if the eigenvalues of $\rho$ majorize the eigenvalues of $\sigma$. In other words,
%\alv{do we want $\bs{\lambda}^{(\rho)} \succ \bs{\lambda}^{(\sigma)}$ or $\rho \succ \sigma$ here?}
\begin{align}
\rho \xmapsto[]{NO} \sigma \iff \rho \succ \sigma \,\;.
\end{align}

In fact, it can be shown that for any Completely Positive Trace Preserving (CPTP) map that has the maximally mixed state as a fixed point $\mathcal{E}(I_d/d)=I_d/d$ (i.e. \emph{unital} maps), the following holds for all quantum states $\rho$:
%\begin{align}
%	\mathcal E  \text{ unital CPTP } \Longrightarrow \rho \succ \mathcal E (\rho) \,\;. \label{unital_CPTP_1}.
%\end{align}
\begin{align}
\mathcal{E}(\rho)=\sigma \Longrightarrow \rho \succ \sigma \,\;.
\end{align}
As NO are also unital maps, the converse also holds: two quantum states $\rho$ and $ \sigma$ with $\sigma$ more mixed than $\rho$, are always connected with a unital quantum channel:
\begin{align}
\rho \succ \sigma \; \Longrightarrow \; \exists \; \mathcal E \text{ unital CPTP with }  \sigma = \mathcal E(\rho )   \label{unital_CPTP_2}
\end{align}
In fact, the two sets of maps coincide for Hilbert space dimensions $d=2$, where unitary interaction with a four-dimensional maximally mixed environment is sufficient \cite{PhysRevA.87.022111,Muller-Hermes2019}. On the other hand, for $d=3$ not all unital quantum maps are noisy operators \cite{haagerup2011factorization}.

%\geo{New, for first referee.} A function $f$ mapping probability vectors to real numbers is called Schur-concave if $f(\bs p) \le f(\bs p')$ for all $\bs p \succ \bs p'$. An important family of such functions are the R\`{e}nyi entropies $S_\alpha (\bs p) \coloneqq \frac{1}{1-\alpha} \log \left(  \sum_i p_i^\alpha \right)$ ($\alpha \ge 0$) \cite{bengtsson2017geometry} \geo{Maybe cite something more specific?}. For instance, the case $\alpha \to 1$ reduces to the usual von Neumann entropy $S_1 (\bs p) = - \sum_i p_i \log(p_i)$ while $S_2 (\bs p) = - \log \left(  P(\bs p) \right)$, where $P(\bs p) \coloneqq \sum_i p_i^2$ is the purity. As a result, in any mixing process (classical or quantum) the von Neumann entropy of the state increases while its purity decreases.

An alternative way for quantifying the mixedness of quantum states is though their R\`{e}nyi entropies. These  are defined as $S_{\alpha}(\rho) \coloneqq \frac{1}{1-\alpha}\log \tr(\rho^\alpha)$ with $\alpha \ge 0$, such that $\alpha \to 1$ reduces to the usual von Neumann entropy $S_1 (\rho) = - \tr( \rho \log(\rho) )$, while $S_2 (\rho) = - \log \tr (\rho^2)$ is directly related to the purity. We note, however, that these entropies provide a weaker notion of mixedness than majorization does, and do not fully capture the mixing power of NO. Although it is straigthforward to show that \cite{gour2015resource}
\begin{align}
 \rho \succ \sigma \Longrightarrow S_{\alpha}(\rho) \le S_{\alpha}(\sigma) \,\, \forall \alpha \,\;,
\end{align}
the converse statement is not true, $ S_{\alpha}(\rho) \le S_{\alpha}(\sigma) \,\, \forall \alpha \not \Rightarrow \rho \succ \sigma $. For counter-examples see \cite{jonathan1999entanglement}.

%In fact, the two sets of maps coincide for Hilbert space dimensions $d=2$ and $d=3$ \cite{haagerup2011factorization}.  We refer the reader to \cite{gour2015resource} for a review of the framework of unital maps and NO.
%\alv{do we want to say here that these are not the sets we care about, or is it clear from introduction?}\geo{I think we should repeat, something like that? I like redundancy :P}\alv{I think I prefer to take sentence out} In what follows, we are going to restrict the accessible unital maps to those arising from a unital Markovian dissipation, combined with coherent control.

\subsection{Classical and quantum Markovian mixing}

Markovian processes continuous in time give rise to one-parameter families of stochastic matrices that obey the semigroup property
\begin{align}
M(t_1 + t_2) &= M(t_1)M(t_2) \\
M(0) &= I
\end{align}
for any $t_1,t_2 \ge 0$. Such families emerge as solutions to the equation $ \frac{d}{dt} M(t) = Q  M(t) $ for some time-independent generating matrix $Q$. The solution $M(t) = \exp \left( t Q \right)$ is a 1-parameter family of stochastic matrices for all $t \ge 0$ if and only if the generator $Q$ satisfies the following two conditions:
\begin{subequations}\label{bistochastic_generator_constraints}
\begin{align}
\left( Q \right)_{ij} &\ge 0 \quad \text{for $i \ne j$} \label{stochastic_generator_constraint_1}\\
\sum_i \left( Q \right)_{ij} &= 0  \quad \text{for every $j$} \,\;. \label{stochastic_generator_constraint_2}
\end{align}

One can impose extra conditions on the generator $Q$ in order to guarantee that the family $M(t)$ is not just stochastic, but also bistochastic.
%i.e., the maximally mixed probability distribution $\tau$ is a fixed point.
This translates to the additional condition
\begin{align} \label{bistochastic_generator_constraint_3}
\sum_j \left( Q \right)_{ij} &= 0  \quad \text{for every $i$} \,\;. 
\end{align}
\end{subequations}
As a result, if conditions \eqref{bistochastic_generator_constraints} simultaneously hold for the generator $Q$ then the resulting semigroup $M(t)$ describes a mixing processes. For example, the matrix
\begin{align}\label{Q_matrix_elements}
Q = 
\begin{pmatrix}
-\gamma & \gamma \\
\gamma & -\gamma
\end{pmatrix}
\end{align}
for $\gamma >0$ generates a family of T-transforms [see Eq.~\eqref{T_transform_example}] with parameter $s(t) = \frac{1}{2} \left( 1 + e^{- 2 \gamma t} \right) $.

Here we consider quantum Markovian master equations, also called dynamical semigroups \cite{breuer2002theory}.
These give rise to 1-parameter families of CPTP maps which again obey the semigroup property
\begin{align}
\mathcal E (t_1 + t_2) &= \mathcal E (t_1) \mathcal E (t_2)\\
\mathcal E (0) &= \mathcal I
\end{align}
for $t_1,t_2 \ge 0$. The corresponding differential equation is $\frac{d}{dt} {\mathcal E}(t) = \mathcal L \, \mathcal E(t)$ with solution $\mathcal E (t) = \mathcal \exp \left( t \mathcal L \right)$. The solution is a CPTP map for all $t \ge 0$ if and only if the generator $\mathcal L$ (referred to as \textit{Lindbladian}) can be cast into the Lindblad diagonal form:
\begin{multline} \label{Lindblad_form}
\mathcal L (X) = - i \left[  H, X \right] \\ + \sum_\alpha \left[ L_\alpha X L_\alpha^\dagger -\frac{1}{2} \left( L_\alpha^\dagger L_\alpha X + X L_\alpha^\dagger L_\alpha \right)  \right] \,\;,
\end{multline}
for some Hermitian operator $H$ (the \textit{effective Hamiltonian}) and a family of operators $\left\{ L_\alpha \right\}_\alpha$ (the \textit{Lindblad operators}), which we assume to be time-independent. Given a particular master equation and an initial state, the set of quantum states that can be reached are only those that belong to the trajectory $\mathcal{E}_t(\rho)$ which is spanned by a single parameter.

%\begin{lemma}
%Let $\mathcal A$ be the complex operator algebra generated by a set of Lindblad operators $\left\{ L_\alpha \right\}_\alpha$ together with the identity. Then, any other set $\left\{ L'_\alpha \right\}$  produced by employing the freedom of equations \eqref{Lindblad_freedom_1} and \eqref{Lindblad_freedom_2} generates the same algebra $\mathcal A$. 
%\end{lemma}

For our considerations, we impose that the dynamics is unital, i.e., $ \mathcal E (t) \left( I \right) = I$ for all $t \ge 0$ which is equivalent to $\mathcal L \left( I  \right) = 0$ or
\begin{align}
	\sum_\alpha L_\alpha ^\dagger L_\alpha = \sum_\alpha L_\alpha  L^\dagger_\alpha \,\;. \label{unital_Lindblad_operators}
\end{align}
This includes a large amount of dissipative processes found in nature, such as dephasing processes, or thermalizations with an environment in the high temperature limit.

\subsection{Coherent control}

In order to enlarge the set of states that can be reached under the present noise models, we also include in the set of operations an arbitrary amount of coherent control. 

More specifically, we assume that the implementable Hamiltonians $H(t)$ span the  $\mathfrak{su}(n)$ algebra. For our purposes, it will be sufficient to consider $H(t) = \sum_{i=1} ^{d^2-1} c_i(t) H_i$ with continuous control functions $c_i(t)$ that are unbounded ($\left\{H_i \right\}_i$ is a basis of the $\mathfrak{su}(n)$ algebra). We suppose that these control pulses are of frequencies high enough that they do not induce a time dependence on the Lindblad operators of the master equation.

%\geo{new} Although a first-principles analysis of the conditions for such an approximation to hold is beyond the scope of this work, we nevertheless note the following. Ignoring the induced time dependence on the Lindblad operators \st{can be in consistence, for example,} \alv{is consistent} with the standard microscopic derivation of the Lindblad master equation in the weak coupling limit, where one invokes the Born-Markov and the rotating-wave approximations (see, e.g., \cite{breuer2002theory} for the details). There, although the dissipative part of the master equation depends on the system Hamiltonian (and hence on the control protocol), in the limit of strong control pulses  with short duration $ \left| c_i(t) \right| , (\Delta T_{\text{pulse}})^{-1}  \gg g^2 \left|  \gamma_{\alpha \beta}(\omega) \right|$ ($ \gamma _{\alpha \beta}$ denotes the hermitian part of the Fourier transforms of the reservoir correlation functions and $g$ the system-reservoir coupling strength) this dependence can be ignored.

%Having this setup in mind, we define state convertibility as follows:

%This setup (or slight variations thereof) has been previously considered in the quantum control literature, see for example \cite{altafini2004coherent,altafini2003controllability,
%	yuan2012reachable,wu2007controllability,schirmer2002criteria,
%	kurniawan2009controllability}, with emphasis on quantum control properties, most notably of \textit{accessibility}, for which tools from the theory of Lie algebras are usually used. \alv{ok to repeat this?}

\section{Conversion of states under Markovian dissipation and coherent control} \label{section_tools}

\subsection{Setting of the problem}\label{sec:setting}

We are now ready to define the state conversion problem.
We say that the state $\rho$ can be converted to the target state $\sigma$ under unitary control and dissipation $\mathcal L_0$, if there exist a 1-parameter family of Hamiltonian operators $H(t)$  such that the time evolution of the state $\frac{d \rho}{d t}= - i \left[   H(t) , \rho \right] + \mathcal L_0 (\rho)$
%\st{$\rho$, governed by the Lindbladian  $\mathcal L(t) = - i \left[   H(t) , (\cdot) \right] + \mathcal L_0 (\cdot) $},
can approximate $\sigma$ arbitrarily well. We shall denote this as
  \begin{gather}
    \conv 
  \end{gather}
(CC above stands for ``Coherent Control'').

%(T) =  \left[ \Texp \left( \int_0 ^T dt \, \mathcal L(t)  \right) \right] \left( \rho \right)$

Without loss of generality, $\mathcal L_0$ above can be assumed to have a vanishing Hamiltonian part as far as state convertibility is concerned. Furthermore, notice that a state $\rho$ is considered convertible to $\sigma$ even if the conversion process requires infinite time or unbounded pulse strength. This limit, as we will argue now, drastically simplifies the analysis.

% by reducing the state convertibility to a sole property of spectrum of the relevant states.
The strength of the external control fields $\left\| H(t) \right\|$ is unbounded and hence allows for ideal (Dirac delta) pulses to be arbitrarily well approximated. On the other hand, the strength of the dissipation is necessarily bounded, since we are considering time independent Lindbladians $\mathcal L_0$ in finite dimensions. Physically, this means that the rate of noise (or the strength of the interaction with the environment) is much smaller than the speed of the coherent control. This separation of time scales
%also noted  in \cite{altafini2003controllability,rooney2018steering},
simplifies the convertibility problem, effectively transforming it to a property of the spectra of the initial and target states (as multisets, i.e., counting multiplicity of eigenvalues).

This follows from the fact that the convertibility $\rho \xmapsto[]{\mathcal L_0 + \text{CC}} \sigma$ also implies that the conversion $\mathcal U \rho \xmapsto[]{\mathcal L_0 + \text{CC}} \mathcal V \sigma$ is possible, where $\mathcal U, \mathcal V$ are any unitary superoperators. %\footnote{This is a well-known fact in the quantum control literature, see, e.g., \cite{altafini2003controllability,rooney2018steering}}.
This is because there exists (by assumption) a sequence of Hamiltonian protocols $H(t)$ such that the time evolution of $\rho$, i.e., $\mathcal E (\rho) \coloneqq\left[ \Texp \left( \int dt \, \mathcal L(t)  \right) \right] \rho $ approximates $\sigma$ arbitrarily well. One can construct a modified protocol by prepending and appending two extra pulses of duration $\delta t$ that generate $\mathcal U ^\dagger$ and $\mathcal V$, respectively. However, in the limit of ideal pulses $\delta t \to 0$ the time evolution becomes $\mathcal E ' = \mathcal V \mathcal E\mathcal U ^\dagger$ \footnote{This claim has appeared in various places in the relevant literature. See, e.g.,  \cite{rooney2018steering} for a more detailed and rigorous proof.}. %Hence, the possibility of conversion $\rho \xmapsto[]{\mathcal L_0 + \text{CC}} \sigma$ for a given Lindbladian $\mathcal L_0$ depends only on the spectra of $\rho$ and $\sigma$ 

%The above fact \alv{assumption rather?}\geo{What about ``fact''? We can discuss} implies that the possibility of conversion $\rho \xmapsto[]{\mathcal L_0 + \text{CC}} \sigma$ for a given Lindbladian $\mathcal L_0$ depends only on the spectra of $\rho$ and $\sigma$. 
In addition, if $\rho_1 \xmapsto[]{\mathcal L_0 + \text{CC}} \rho_2$ and $\rho_2 \xmapsto[]{\mathcal L_0 + \text{CC}} \rho_3$ hold true, then we can deduce $\rho_1 \xmapsto[]{\mathcal L_0 + \text{CC}} \rho_3$ (by concatenating the two protocols), i.e., \textit{transitivity} holds. Thus, our set of operations induces again a preorder in the set of quantum states.

For any unital Lindbladian $\mathcal L_0$ we clearly have
\begin{align}
\conv \; \Longrightarrow \; \rho \succ \sigma \,\;,
\end{align}
since the time evolution is unital. However, given any two states such that one is more mixed that the other, is it possible to guarantee that
\begin{align} \label{main_question}
\rho \succ \sigma  \; \xRightarrow{\;\text{ ? } \;}   \conv \; \,\;,
\end{align}
for all such states? That is, is the preorder induced by $\mathcal L_0$ also the majorization preorder as in NO? 

%Clearly, the answer to the above question is a function of the type of dissipation $\mathcal L_0$.
%\alv{we could get rid of this ``trivial remark"?}For example, in the trivial case of controlled \textit{closed} quantum dynamics, $\rho \xmapsto[]{\text{CC}} \sigma$\alv{got rid of $0+ CC$} if and only if the two states have the same spectra (with multiplicities taken into account).

%\alv{I'd delete this as its sort of obvious: hence the answer to the above question in that case is negative: for the task of converting a state $\rho$ to the more mixed state $\sigma$, noise is as a resource.}

\subsection{Evolution equation for the spectrum} \label{subsection_Evolution_equation_for_the_spectrum}

%Motivated by the discussion of the previous section, w
We now seek a differential equation that describes the time evolution of the spectrum of a state under Markovian dynamics, by examining the time evolution of $\rho$ from an orthonormal eigenbasis $\{ \ket{i(t)} \}_i$. Let us assume for now that the choice is such that the eigenvectors are differentiable with respect to time. This assumption might fail in general at points where the spectrum of $\rho$ is degenerate, but this poses no problem for our construction.
%We will comment on that later.\alv{maybe just comment later instead of announcing it}

Let $\dot \rho(t) = -i \left[H(t), \rho(t) \right] + \mathcal L_0 $ be the evolution equation for the state $\rho(t)$, under some  Hamiltonian protocol $H(t)$. Then, the corresponding evolution equation for the eigenvalues $\lambda_i(\rho) \coloneqq \braket{i(t) |\rho(t) | i(t)} $ is
  \begin{gather} \label{eigval_diff_eq}
    \dot \lambda_i = \sum_j \left[ Q(t) \right]_{ij} \lambda_j \,\;,
  \end{gather}
  where
  \begin{subequations} \label{Q_matrix_all}
  \begin{gather} \label{Q_matrix_def_eq}
    \left[Q(X)\right]_{ij} \coloneqq   X_{ij} - \delta_{ij} \sum_k X_{kj}    \,\;, \\
    X_{ij} \coloneqq \sum_\alpha X_{ij}^{(\alpha)} \,\;,\\
    X^{(\alpha)}_{ij} = \left| \braket{i(t) | L^{(\alpha)} | j(t) } \right|^2  \,\;.  \label{X_matrix_def}
  \end{gather}
	\end{subequations}  
  
This can be shown as follows. We have $\dot \lambda_i = \braket{i| \dot \rho | i}$ (since $\braket{i|i} = 1$). By inserting the Lindblad form [Eq.~\eqref{Lindblad_form}] for the dissipative part $\mathcal L _0$ and using the spectral decomposition of $\rho$ we get, as desired,
	\begin{align*}
		\dot \lambda_i &= \sum_\alpha \left( \sum_j X_{ij}^{(\alpha)} \lambda_j - \sum_k X_{ki}^{(\alpha)} \lambda_i \right) \\
		& = \sum_j \left(  \sum_{\alpha} \left[ X_{ij}^{(\alpha)} - \sum_k X_{kj}^{(\alpha)} \delta_{ij} \right] \right) \lambda_j \\
		& = \sum_j Q_{ij} \lambda_j \,\;.
	\end{align*}

By construction, the matrix $Q$ is a generator of stochastic matrices, namely Equations \eqref{stochastic_generator_constraint_1} and \eqref{stochastic_generator_constraint_2} are satisfied for all $Q$ matrices arising as in equations \eqref{Q_matrix_all}, for any set of Lindblad operators. If $\mathcal L_0$ is in addition unital, Equation \eqref{bistochastic_generator_constraint_3} is also satisfied, as it can be easily checked directly invoking Eq.~\eqref{unital_Lindblad_operators}.
Notice that $Q$ depends jointly on the dissipative and the Hamiltonian part of $\mathcal L$. The dissipative part $\mathcal L_0$ directly determines the $Q$ matrix through the set of operators $\{L^{(\alpha)}\}_\alpha$, while the Hamiltonian part influences the time evolution of the eigenbasis $\left\{ \ket{i(t)} \right\}_{i}$.

Equation \eqref{eigval_diff_eq} describes the evolution of the eigenvalues of $\rho(t)$ under Lindblad evolution. Perfect coherent control of the quantum system allows steering the eigenbasis of $\rho(t)$ (arbitrarily close) to any desired instantaneous orthonormal eigenbasis. Hence, the eigenbasis can be regarded as the control parameter of the system, in which case Eq.~\eqref{eigval_diff_eq} describes the corresponding evolution of the eigenvalues.

More specifically, for any choice of a piecewise differentiable orthonormal basis $\bs B = \left\{ \ket{i(t)} \right\}$ with a finite number of discontinuities, there exists a Hamiltonian protocol $H(t)$ such that the eigenbasis of $\rho(t)$, evolving  under $\mathcal L = \mathcal K_{H(t)} + \mathcal L_0$ (where $\mathcal K_H (X) \coloneqq -i \left[ H , X \right]$), arbitrarily well approximate the prescribed basis $\bs B$ \footnote{Note that the initial condition $\rho(0)$ is free, i.e., we do not need to require a compatibility condition to guarantee that the eigenvectors of $\rho(0)$ can be chosen to coincide with $\bs B$ for $t=0$. This is because the eigenbasis of the resulting evolution $\rho(t)$ \textit{arbitrarily well approximates} $\bs B$ and it does not necessarily coincide with it for all $t$ of the protocol.}. This fact was proven by Rooney \textit{et al.} in Ref.~\cite{rooney2018steering}, where an explicit expression for $H(t)$ (as a function of $\bs B$ and $\mathcal L_0$) was given. We will invoke this fact later on by specifying as input to specific evolutions the eigenbasis $\bs B$ instead of the Hamiltonian control protocol. Notice that for any such prescribed eigenbasis $\bs B$ the corresponding evolution equation of the eigenvalues of $\rho(t)$ is given by Eq.~\eqref{eigval_diff_eq}, where the input regarding $\bs B$ is incorporated in Eq.~\eqref{X_matrix_def}.

%\st{Equation} \eqref{eigval_diff_eq} \st{is our main tool for analyzing the different properties of mixing Lindbladians. We use it in the following section to construct protocols that achieve the desired mixing.} \alv{I think this is redundant} 

Before proceeding, we comment on a relevant technical point. The eigenbasis of a non-degenerate density operator is uniquely defined, but the same cannot be said when degeneracy is present. Differentiability of $\rho(t)$ (automatically satisfied by Markovian evolutions) guarantees that its eigenbasis $\left\{ \ket{i(t)} \right\}$ is differentiable at regions of non-degeneracy. However, differentiability of $\rho(t)$ alone does not guarantee the existence of a differentiable eigenbasis at points (or regions) where degeneracy is present. Hence, one might worry that the evolution equation \eqref{eigval_diff_eq} might not apply directly in such cases. Furthermore, it is unclear what is the relevant eigenbasis for Eq.~\eqref{X_matrix_def}.

Let us, nevertheless, note the following.
\begin{inparaenum}[(i)]
\item The eigenvalues of the density operator can always be chosen to be differentiable in any interval if $\rho(t)$ is itself continuously differentiable (i.e., it's derivative exists and is continuous) \cite{kato2013perturbation}, which is true in our case. Furthermore, \item the set of degenerate states is a zero-measure set in the state space. This implies that, given a degenerate state $\rho$, one can always consider a non-degenerate counterpart $\tilde \rho$ that is as close as desired to the original $\rho$. Then, by continuity arguments, the evolutions of $\rho$ and $\tilde \rho$ under some common $\mathcal L(t)$ are in practice indistinguishable. On more physical grounds, states in the laboratory are never exactly degenerate.
\end{inparaenum}

This suggests that nothing exceptional happens with the evolution of degenerate states, but the possible non-differentiability of the eigenvectors at crossing points is an artifact of our choice to work with one-dimensional projectors, instead with the total projector in the degenerate subspace (which is differentiable).

\subsection{A first example: Depolarizing noise}

Let us show a simple example of a Master equation with which transitions are very limited. This also demonstrates how certain kinds of Markovian noise heavily restrict the preorder of states. Consider the \textit{depolarizing} Lindbladian in $d$ dimensions
	\begin{align} \label{depolarizing_Lindbladian}
	\mathcal L_0 (X) = \frac{I_d}{d} \Tr \left( X \right) - X \,\;.
	\end{align}
 $\mathcal L_0$ is unital and the corresponding Lindblad operators $\left\{ L_\alpha \right\}_\alpha $ can all be chosen to be unitary \footnote{In fact, the Lindblad operators can also be thought of as a unitary 1-design.}.

In absence of any control fields, the time evolution due to the Lindbladian above has the simple form
\begin{align}
	\rho(t) = e^{-t} \rho(0) +  \left( 1 - e^{-t} \right) \frac{I}{d}  \,\;.
\end{align}
Thus the spectrum of the noised state is a mixing between the original probability distribution and the maximally mixed one, namely
\begin{align}
\lambda_i(t) =  e^{-t} \lambda_i(0) +  \left( 1 - e^{-t} \right) \frac{1}{d} \,\;.  \label{evolution_eigenvalues_depolarizing}
\end{align} 

The highly symmetric form of the depolarizing Lindbladian has the additional property that the evolution of the eigenvalues is independent of any external coherent control. In other words, $\conv$ for depolarizing noise if and only if
\begin{align}\label{eq:idmixing}
\bs{\lambda}(\sigma) = s \bs{\lambda}(\rho) + (1-s) \bs{\lambda}(I_d/d)
\end{align}	
for some $s \in \left[ 0 , 1 \right]$. 
This is because the depolarizing Lindbladian $\mathcal L_0$ commutes with any Hamiltonian part
\begin{align}
	\left[ \mathcal L_0 , \mathcal K_H \right] = 0 \,\;,
\end{align}
which follows from $\mathcal L_0 \mathcal K_H = -\mathcal K_H$ and $\mathcal K_H \mathcal L_0 = -\mathcal K_H$. As a result, for any protocol $H(t)$ ($t \in \left[ 0 , T \right]$) the propagator can be split as
\begin{align}
\Texp  \left( \int_0 ^T dt \, \left[ \mathcal K_{H(t)} + \mathcal L_0 \right] \right) = \mathcal U \exp \left( T \mathcal L_0 \right) \,\;,
\end{align}
with $\mathcal U =  \Texp  \left( \int_0 ^T dt \,  \mathcal K_{H(t)} \right) $. However, the action of $\mathcal U$ on the a state does not affect its eigenvalues, hence the eigenvalue evolution equation is identical to Eq.~\eqref{evolution_eigenvalues_depolarizing} as in the original system (with the absence of control).

The above demonstrates the limited value of depolarizing noise for mixing tasks in $d > 2$. For qubits, the preorder is specified by a single parameter, thus any mixing evolution of the eigenvalues is of the form \eqref{eq:idmixing}, so in fact any kind of unital noise is sufficient for Eq.~\eqref{main_question} to hold \footnote{The case $d=2$ is trivial due to the elementary fact that the probability simplex is 1-dimensional.}.

\section{Lindbladians with optimal mixing properties} \label{section_main}

We now characterize the set of Markovian master equations that allow us to recover the majorization preorder. The relevant definition is as follows:
\\ \\
\textit{A unital Lindbladian $\mathcal L_0$ is  \textbf{optimal} if and only if $\conv$ for any pair of states satisfying $\rho \succ \sigma$.}
\\ \\
First, we provide the general statement that characterizes the whole set of optimal Lindbladians. Then, we focus on a physically relevant subset of such operations, namely the class of dephasing maps. For the dephasing case we give an alternative construction that demonstrates their optimality, based on the Schur-Horn theorem \cite{bhatia2013matrix}.

\subsection{Optimal Lindbladians can mix exactly two levels at a time}
We now show the following:
\\ \\
\textit{A unital lindbladian $\mathcal L_0 \neq 0$ is optimal if and only if there exists an ordered orthonormal basis $\bs B'$ in which all the corresponding Lindblad operators $\left\{ L_\alpha \right\}_\alpha$ admit a matrix representation of the form
\begin{align}
L_\alpha = M_\alpha \oplus D_\alpha \quad \text{$\forall$ $\alpha$}\,\;, \label{Lindblad_decomposition_optimal}
\end{align}
\\
%$L_\alpha = M_\alpha \oplus D_\alpha$
where each $M_\alpha \in \mathbb M_2 (\mathbb C)$ is a $2 \times 2$ block and $D_\alpha \in \mathbb M_{d-2} (\mathbb C)$ is a diagonal matrix.}
\\ \\	
Let us consider a pair of states satisfying $\rho \succ \sigma$. By the majorization assumption, it follows that there exists a bistochastic matrix $B_{ij}$ and a series of $T$-transforms such that 
\begin{gather}
\lambda_i(\sigma) = \sum_j B_{ij} \lambda_j(\rho) \\
B = T_{(i_kj_k)}(s_k) \cdot T_{(i_{k-1} j_{k-1})}(s_{k-1}) \cdot \dotso \cdot T_{(i_1 j_1)}(s_1) \,\;, \label{tdecomp}
\end{gather}
with $k \le d-1$. Each $T$-transform is of the form 
\begin{align} 
T_{(ij)}(s) = (1-s) I + s P _{(ij)} \,\;,
\end{align}
where $P_{(ij)}$ is the transposition of the $(ij)$ levels and $s \in \left[ 0 , 1\right]$. For reasons that will become clear, we want to restrict $s \in \left[ 0 , 1/2 \right]$, for which we utilize the relation
\begin{align}
T_{(ij)}(1-s) = P_{(ij)} T_{(ij)}(s) 
\end{align}
and alter all T-transforms with $s > 1/2$, at the expense of inserting the required transposition matrices in the decomposition of $B$. The occurring permutation matrices can be brought to the rightmost of $B$ by using the relations
\begin{align}
P_{(ij)} T_{(ik)} = T_{(jk)} P_{(ij)} \;, \quad i,j \ne k \,\;.
\end{align}
This results in a decomposition of $B$ similar to that of Eq.~\eqref{tdecomp} but with $s_i \in \left[0,1/2 \right] $ and possibly some permutation matrix on the rightmost side, which we will not write explicitly as it will turn out to be unimportant.
%\st{From now on we always assume that any $B$ is decomposed in this way.}

The next step is to break the conversion problem into $k$ pieces, each of them corresponding to one of the T-transforms in the decomposition of $B$. We have already argued that
\begin{inparaenum}[(i)]
\item the convertibility (or impossibility thereof) is a property of the eigenvalues of the initial and target states (since one can implement fast unitaries at the beginning and in the end), and that \item the state conversion is transitive, i.e., we can break the total conversion into intermediate steps (if each of them is possible then the total transformation is also possible).
\end{inparaenum} 
It hence follows that all conversions $\conv$ (with $\rho \succ \sigma$) are possible if
%there exist coherent control protocols that result in mixing of the eigenvalues of the state as described by the family
the family of T-transforms $T_{(ij)}(s)$ with $s \in \left[0,1/2 \right]$ (for all pairs $(ij)$) are implementable.

Notice, however, that if a transformation $T_{(ij)}(s)$ is implementable then also any other $T_{(i'j')}(s)$ is, just by exchanging the populations $i \leftrightarrow i'$, $j \leftrightarrow j'$ (which is a unitary transformation), implementing $T_{(ij)}(s)$ and finally exchanging them back.
In addition, the permutation matrix from the decomposition of $B$ mentioned earlier can be considered as part of a initial unitary that is potentially needed in the beginning of the protocol.

It remains to show that the transformations $T_{(ij)}(s)$ (for all $s \le 1/2$ and for some pair of levels $(ij)$) are implementable. For that, we are going to invoke Eq.~\eqref{eigval_diff_eq} together with the main assumption that the Lindblad operators can be cast into the form \eqref{Lindblad_decomposition_optimal}.
We will treat the case $s = 1/2$ separately, so we assume $s < 1/2$ for now. 

For convenience, and without loss of generality, we consider an initial state $\rho_1$ and an (intermediate) target state $\rho_2$, with $\lambda(\rho_2) = T_{(ij)}(s) \lambda(\rho_1)$, which are both diagonal in the $\bs B'$ basis.
%\st{(this poses no loss of generality since fast pulses can be applied to achieve that)}. \st{For our construction, we need to}
Let us specify the eigenbasis $\bs B$ in Eq.~\eqref{X_matrix_def} (which we treat as the control parameter, as explained earlier), for which we will distinguish two different cases: $(i)$ at least one $M_\alpha$ is non-diagonal in $\bs B'$, $(ii)$ all $M_\alpha$'s are diagonal in $\bs B'$.

$(i)$ We choose a basis $\bs B$ coinciding with $\bs B'$ such that the levels $(ij)$ %\st{(that we intend to be affected by the T-transform)}
correspond to the $2 \times 2$ blocks $M_\alpha$. Now we invoke Eq.~\eqref{eigval_diff_eq} to read what will be the time evolution of the eigenvalues. The corresponding matrix $Q$ has the form
\begin{align}
Q = Q_{(ij)} \oplus 0 
\end{align}
where $Q_{(ij)}$ acts on the $(ij)$ levels. By the unitality assumption of $\mathcal L_0$, the matrix $Q$ is a valid generator of  bistochastic matrices, so it automatically satisfies all constraints given by Eqs.~\eqref{bistochastic_generator_constraints}. Hence $Q_{(ij)}$ is necessarily of the form \eqref{Q_matrix_elements}, i.e.,
\begin{align*}
Q_{(ij)} = \begin{pmatrix}
	-\gamma  & \gamma \\
	\gamma & -\gamma 
	\end{pmatrix}  \,\;.
\end{align*}
In addition, at least one $M_\alpha$ is non-diagonal by assumption, so from Eqs.~\eqref{Q_matrix_all} we get $\gamma = \sum_\alpha \left|  \left(M_\alpha\right)_{12} \right|^2 > 0$.
%Finally, there exists a driving protocol such that this $Q$ remains constant in time. 
By maintaining the prescribed $\bs B$ constant in time, the
%\st{exponential of $Q$ (which governs the evolution of the eigenvalues) is}
T-transform is
\begin{subequations} \label{Q_solution}
\begin{gather}
\exp\left( Q t \right) = T_{(ij)}[s(t)] \\
s(t) = \frac{1}{2} \left( 1 - e^{- 2 \gamma t} \right) \,\;, \label{s_time_depend}
\end{gather}
\end{subequations}
hence the desired family of T-transforms can be implemented.

$(ii)$ First, notice that the previous choice of coinciding bases $\bs B$ and $\bs B'$ is not adequate to generate T-transforms, since now $\gamma = 0$. Nevertheless, a slightly different choice of the driving basis $\bs B$ can fix this problem. 

The assumption of non-vanishing dissipation $\mathcal L_0 \ne 0$ implies that there exist at least two diagonal elements $(ij)$ in the representation \eqref{Lindblad_decomposition_optimal} such that $\left( L^{(\alpha)} \right)_{ii} \ne \left( L^{(\alpha)} \right)_{jj}$ for some $\alpha$. Indeed, $L_\alpha \propto I$ for all $\alpha$ implies $\mathcal L_0 = 0$, which is excluded. Therefore we consider $\bs B$ that coincides with $\bs B'$ except on the $(ij)$ levels that the two bases are connected via a Hadamard unitary rotation, namely
%A possible choice \footnote{In general, $U$ may be time dependent, i.e., $\bs B$ can be time dependent.Here we make a choice such that there is no time dependence for as long as a single T-transform is being implemented.} is the unitary matrix
%that acts as a Hadamard (w.r.t. the computational basis) between the $(ij)$ levels and trivially on the rest, i.e.,
$U = U_H \oplus I$, where
\begin{align}
U_H \coloneqq \frac{1}{\sqrt{2}} \begin{pmatrix}
1 & 1\\
1 & -1
\end{pmatrix} \,\;.
\end{align}
Proceeding like in the previous case, we can calculate the corresponding $Q$ matrix.
% for the time interval that the eigenbasis of $\rho$ is $\bs B$.
By use of Eqs.~\eqref{Q_matrix_all}, we again get $Q = Q_{(ij)} \oplus 0$,
where now $\gamma = (1/4) \sum_{\alpha} \left| \left( L^{(\alpha)} \right)_{ii} - \left( L^{(\alpha)} \right)_{jj} \right|^2$. The exponential of $Q$ has again the desired form \eqref{Q_solution}. As we argued, there always exist levels such that $\gamma > 0$.  Hence, we again conclude T-transform for $s \in \left[ 0 , 1/2 \right)$ can be implemented.

It remains to comment on the $s = 1/2$ case. Although the above construction would require an infinite time to achieve $s=1/2$, this poses no problem since by allowing larger and larger total time $t$ for the protocol we can approximate the $s=1/2$ T-transform arbitrarily well.

%As a result, we can satisfy the conditions of Lemma~\autoref{interconversion_lemma}, by choosing $T$ such that $s(T) = s$ in Eq.~\eqref{s_time_depend}, $U(t)$ (constant) as described  and $q_i = \lambda_i(\rho_2)$, which works by construction.

%Notice that this includes dephasing Lindbladians as a special case, i.e., when the $M_\alpha$'s admit a diagonal representation in $\bs B$. Also, once again, the freedom prescribed by equations \eqref{Lindblad_freedom_1} and \eqref{Lindblad_freedom_2} for the different representations in Lindblad operation does not alter the existence (or lack thereof) of a representation of the form \eqref{Lindblad_decomposition_optimal}. 

%Let us begin by proving sufficiency. The arguments are just a variant of the arguments we already provided for dephasing case. As we already explained there, it is enough to show that any T-transform $T_{(ij)}(s)$ for $s \in  \left[ 0 , 1/2 \right)$ on the eigenvalues is implementable.

Now we proceed to show the necessity of the form \eqref{Lindblad_decomposition_optimal}.
%\st{of the Lindblad operators is also necessary}.
We consider a system with Hilbert space dimension $d \ge 3$, otherwise the aforementioned form is always attained. Let us assume that the form  \eqref{Lindblad_decomposition_optimal} is not admittable (in any orthonormal basis).
%\st{Then, there exist target states $\sigma$ whose spectrum differs just by a 2-level mixing from the initial state $\rho$ that cannot be reached.}
%\st{Let us elaborate further.}
The non-existence of an orthonormal basis such that the Lindblad operators can be all brought to a form of Eq.~\eqref{Lindblad_decomposition_optimal} implies that the $Q$ matrix (Eqs.~\eqref{Q_matrix_all}) is always mixing at least three levels at all times.
%\st{On the other hand, mixing is an irreversible process. Hence, one can}
If we consider an initial state $\rho$ and a target state $\sigma$, with $\rho \succ \sigma $, such that the spectrum of $\sigma$ differs solely by a two-level mixing, it is clear that approximating $\sigma$ arbitrarily well is impossible.

Notice that, although the driving basis $\bs B$ above was chosen (in both cases $(i)$ and $(ii)$) to be time-independent during the implementation of a single T-transform, in general this does not imply that coherent manipulations are absent during that time interval.

\subsection{All Dephasing Lindbladians are optimal}

We now focus on a particular physically relevant case of the above: dephasing master equations. We say that a time-independent Lindbladian $\mathcal L \ne 0$ \textit{dephases} if
%as for $t \to \infty$ \st{the corresponding time is dephasing, i.e.,} we have that
\begin{align} \label{dephasing_projectors}
	\lim_{t \to \infty} \exp \left(\mathcal L t \right)  = \sum_i \Pi_i \left( \cdot \right) \Pi_i
\end{align}
for some complete family of orthogonal projectors $\left\{ \Pi_i \right\}$ (not necessarily rank-1).
%The diagonalizability condition for the Lindblad operators  is equivalent to the condition that Lindbladian $\mathcal L_0 \ne 0$ dephases.
Dephasing Lindbladians admit a representation in Lindblad operators such that all $ L^{(\alpha)}$ are simultaneously diagonalizable. As a result, $M_\alpha$ in Eq. \eqref{Lindblad_decomposition_optimal} are also diagonal for all $\alpha$. Notice that dephasing Lindbladians are unital. Hence we have the following:
\\ \\
\textit{
Any dephasing Lindbladian is optimal.}
\\ \\
This can also be shown by using the Schur-Horn theorem \cite{marshall2010inequalities}, which provides an alternative protocol for state conversion.

Assume $\rho \succ \sigma$. Then, the Schur-Horn theorem guarantees the existence of a unitary transformation $U$ such that $\bs{\lambda} (\sigma)  = \bs{\lambda} ( \mathcal D \left(U \rho U^\dagger \right))$, where $\mathcal D (X) \coloneqq \sum_{i=1}^d P_i X P_i $ is any dephasing channel with $ \left\{P_i \coloneqq \ket{i}\bra{i} \right\}_i$ a complete family of rank-1 projectors. In the case that all $\Pi_i$'s from Eq.\eqref{dephasing_projectors} are rank-1, it suffices to unitarily rotate the state $\rho$ and then allow the Markovian dissipation to completely dephase the rotated state (without any additional coherent control). The resulting state is unitarily equivalent to $\sigma$. Thus $\conv$ is achievable.

If any of the $\Pi_i$'s have rank greater than one, the resulting state is block-diagonal with respect to the $\left\{\Pi_i \right\}_i$ decomposition and is not unitarily equivalent to the state $\sigma$. Nevertheless, by the Schur-Horn theorem, there exists a basis such that the eigenvalues of $\sigma$ lie in the diagonal of the dephased state. The off-diagonal elements  surviving after the initial dephasing process can be eliminated by exchanging the populations between different blocks and then dephasing again. By repeating this process one arrives to a diagonal state, unitarily equivalent to $\sigma$. In equations, for any $\ket{k}\bra{l}$ ($k \ne l$) such that $\Pi_i \left(\ket{k}\bra{l}\right) \Pi_i = \ket{k}\bra{l}$ there exists a permutation $\pi \in \mathcal S_d$ such that $\Pi_i \left(\ket{\pi(k)}\bra{\pi(l)}\right) \Pi_i = 0$. Notice that the values of the diagonal elements remain unchanged during this process. %\alv{I don't think repeating this is necessary} \geo{If the word argument is clear, it's not. But someone (e.g., a referee) might think we're waving our hands, while we are not: the argument is precise. Thus I vote to keep it. If you don't like the phrasing change it, but we need some math in my opinion.}

\section{Discussion and outlook} \label{section_results}

%In this work, we made progress on the question of convertibility of quantum states under dynamics dictated by a unital, time-independent Lindbladian together with coherent unitary control. The time evolution generated by this type of Markovian master equations has the defining characteristic of \textit{mixing}, i.e., the spectrum of the output state is always more uniform than the spectrum of the input one. We further identified the conditions for the dissipation $\mathcal L_0$ such that, under unitary control of the system alone, the conversion $\conv$ is always possible for all $\rho$ and $\sigma$ with $\rho \succ \sigma $.

We have characterized the set of Markovian master equations which are optimal for noisy state transformations: dephasing Lindbladians have optimal converting properties, but not all such optimal Lindbladians are dephasing. Together with unitary control of the system alone, they have a converting power which is equivalent to the one of Noisy Operations (NO).

Recently, Boes \textit{et al.} in \cite{boes2018catalytic2} have explored another physical restriction similar in spirit: they show that the smallest possible environment one can have in the resource theory of NO such that the majorization preorder is preserved is of dimension $\lceil d^{1/2} \rceil$. That is, in the resource theory of NO one can interact with an arbitrarily large environment, but the state transitions allowed are the same when there is access to one with the aforementioned size.

The existence of individual master equations that allow for the full preorder of NO to be preserved is very much in contrast to what happens in \emph{thermal operations} (TO). This set includes arbitrary energy-preserving interactions with a finite-temperature bath of any size and Hamiltonian. In that case, the relevant preorder is given by \emph{thermomajorization} \cite{horodecki2013fundamental}, which can be understood as a ``finite temperature" equivalent of majorization in which the Gibbs distribution plays the role of the maximally mixed distribution. It can be shown \cite{lostaglio2018elementary} that in that context, Markovian master equations with thermal states as steady states are far from enough to achieve the preorder of TO, already for single qubits. This is still the case when one allows for full unitary control, as can be seen in \cite{alhambra2018heat}. There, it is shown that non-Markovian interactions with the environment are necessary for certain TO that are optimal in the task of heat-bath algorithmic cooling of individual qubits.

When exploring the physical meaning of resource-theoretic constraints, a natural, complementary question arises: Given a set of further physical restrictions on the operations, what is the preorder induced on the space of states?
Previous work has studied this by placing restrictions on the size \cite{richens2018finite}  or homogeneity \cite{shu2018violation} of the environment, or on the system-environment interactions allowed \cite{wilming2016second,lostaglio2018elementary}. While this has not been the focus of the present work we note that, for NO, limited coherent control may be another such relevant restriction.

Regarding that limitation, the discussion in Sec. \ref{sec:setting} above shows that, at the very least, there are be some states in the orbit $\mathcal{U}(\rho)$ that are out of reach, so the majorization preorder breaks down and the eigenvalues of the state cease to give sufficient information. Thus, in general, the set of states that can be reached will depend non-trivially on the eigenbasis of the initial state, and on the particular relation between the noise model and the amount of control, as well as their relative strengths. We do not expect that there exists a simple answer to this general problem in quantum control theory (for particular settings see, for example, \cite{jirari2005optimal,sauer2013optimal}).

Understanding how different sets of maps act on quantum states is in itself a question of controllability of quantum systems, and can thus find a number of practical applications. 
For instance, dephasing noise has been found to enhance transport in disordered systems. In  \cite{plenio2008dephasing,rebentrost2009environment} this fact is derived assuming that coherent control is limited to the free evolution of a disordered Hamiltonian, and in \cite{cifuentes2017energy} a similar effect is found to occur when one allows for larger coherent control. On top of that, Markovian noise appears as a resource in certain models of computation/quantum simulation \cite{verstraete2009quantum} and its control is the key ingredient of dissipative engineering \cite{diehl2008quantum,pastawski2011quantum,krauter2011entanglement}. These illustrate how in certain situations noise can indeed be seen as an aid rather than a drawback. We hope that the present work will contribute to the solution of the very practical problem of understanding how to utilize noise in quantum settings.

%\alv{I don't quite see this, you could use unitaries to permute eigenstates?}
%Recently, P.~Boes \textit{et al.} in \cite{boes2018catalytic} came up with a protocol for the conversion $\rho  \mapsto \sigma$ (for any $\rho \succ \sigma$) that involves a noisy operation with a maximally mixed ancillary system of size at most $\lceil d^{1/2} \rceil$. The protocol is based on the implementation of a dephasing map and the use of the Schur-Horn theorem. One could follow a similar strategy to show that a dephasing Lindbladian is optimal. Notice, however, that such a construction would only work for total dephasing, i.e., only for rank-1 projectors in Eq.~\eqref{dephasing_projectors}, hence the T-transform construction we have is more general (though more involved).
%Also, the implementation would always require infinite time to implement exactly.

%In conclusion, in the present work we have demonstrated the inequivalent resourcefulness of Markovian dissipations for mixing purposes. So far we have only tackled questions regarding the converting-power for different kinds of dissipations. \alv{I don't understand this}It remains as an open question how the set of \textit{quantum operations} is related to, e.g., noisy operations and unital CPTP maps in the case of optimal dissipation. Furthermore, a meaningful qualitative ranking of different (not only optimal) kinds of dissipation is still missing. These questions, among other, represent possible challenges for future investigation.

\acknowledgments

G.S. is thankful to L.~Campos Venuti for critical reading of the manuscript, N.A.~Rodr\'iguez-Briones for helpful discussions, R.~Di Felice and the CNR-NANO Institute in Modena, Italy for their kind hospitality and acknowledges financial support from a University of Southern Californian Gold Family Graduate Fellowship. P.Z. acknowledges partial support from the NSF award PHY-1819189. This research was supported in part by Perimeter Institute for Theoretical Physics. Research at Perimeter Institute is supported by the Government of Canada through the Department of Innovation, Science and Economic Development and by the Province of Ontario through the Ministry of Research, Innovation and Science.

\bibliography{refs,References}

\end{document}